\documentstyle[12pt]{article}

\newcommand{\beq}{\begin{equation}}
\newcommand{\eeq}{\end{equation}}
\newcommand{\beqn}{\begin{eqnarray}}
\newcommand{\eeqn}{\end{eqnarray}}
\newcommand{\bi}{\bibitem}
\newcommand{\n}{\newline}
\newcommand{\nn}{\nonumber\\}

\newcommand{\bc}{\begin{center}}
\newcommand{\ec}{\end{center}}
\newcommand{\cl}{{\cal C}}
\newcommand{\gl}{{\cal G}}

\begin{document}

\pagestyle{empty}
\bc
{\Large {
{\bf WEAK-ERGODICITY BREAKING IN }
}}
\ec
\bc
{\Large {
{\bf MEAN-FIELD SPIN-GLASS MODELS}
}}
\ec
\vspace{1cm}
\bc
{\large {
L. F. Cugliandolo and J. Kurchan
}}
\ec
\vspace{.5cm}
\bc
{\large {
Dipartimento di Fisica, Universit\`a di Roma I,
{\it La Sapienza},
}}
\ec
\bc
{\large {
I-00185 Roma,
}}
\ec
\bc
{\large {
Italy
}}
\ec
\bc
{\large {
INFN Sezione di Roma I, Roma, Italy
}}
\ec

\vspace{7cm}

\noindent
{\bf Talk given at} \n
{\bf Fifth International Workshop on Disordered Systems } \n
{\bf Andalo Trento } \n
{\bf Italy} \n
{\bf February, 1994}

\newpage
\pagestyle{plain}
\setcounter{page}{1}

The past decade has witnessed a great deal of work on the theoretical
modeling of spin-glasses. The efforts have been
mainly concentrated on the study of the equilibrium properties
of mean-field and low-dimensional models.

Thanks to the  the replica method
there is at present a good understanding of the Gibbs-Boltzmann measure
in  mean-field models
\cite{Mepavi}. The phase-space turns out to have a very complicated landscape
with many hierarchically organised states. These states are
separated by barriers diverging with the
size of the system and hence implying ergodicity breaking.

The implementation of the replica approach for low-dimensional models
has proven to be a hard problem and few  analytical results are at present
available for them.
Whether the picture found from the replica
analysis of mean-field models can be applied to  realistic
systems is still an open question;
the nature of their  ordered state and phase transition remain
controversial \cite{Fihu,Mapari}.
Indeed,  it is sometimes argued that
mean-field models are too unrealistic
and cannot be accepted as candidates to describe
real spin glasses.

However, the experimental evidence suggests that spin-glass
physics in the low temperature phase is essentially dynamical:
they continue to evolve long after thermalization at a
subcritical temperature and their  dynamical properties depend on their history
throughout the experimental time.
These effects are known as {\it aging phenomena}
and are widely observed in many other disordered systems.
Evidently, the  relevance of a model consists in its ability to
capture these phenomena.

Besides, the experiments suggest  that the dynamics of spin-glasses
has similar features to those found in usual glasses \cite{St}.
One may then
hope to apply the developments in the (perhaps simpler) area of
spin-glasses to other glassy systems.

Though several phenomenological approaches to
aging phenomena in spin-glasses have been made \cite{pheno},
a fully microscopic description is still lacking.
It was recently noticed that mean-field models, though schematic, do
possess a rich dynamical phenomenology resembling, at least qualitatively,
the experimental observations \cite{Cuku}.
Furthermore, it turns out that one can obtain analytical results for
their long-time dynamics.

In what follows we shall describe
the long-time dynamical behaviour of two representative examples:
the $p$-spin spherical model defined  by
\beq
H_J[s] = -\sum_{i_1 < \dots < i_p}^N J_{i_1 \dots i_p} s_{i_1} \dots s_{i_p}
+ z(t) \, \left( \sum_{i=1}^N s_i^2 - N \right)
\; ,
\eeq
and the Sherrington-Kirkpatrick (soft-spin) model
\beq
H_J[s] = -\sum_{i_1 < i_j}^N J_{ij} s_i s_j
+ u \, \left( \sum_{i=1}^N (s_i^2 - 1)^2 \right)
\; .
\eeq
The last terms enforce the spin measure (spherically constrained  and
soft spin, respectively),
and the quenched disorder $J$ has in both cases zero mean and variance
$\overline J^2 = p!/ (4 N^{(p-1)})$ and
$\overline J^2 = 1/ (2N)$, respectively.

The analytical study of the long-time relaxation of these models
reveals that both models have aging phenomena, but
with rather different characteristics \cite{Cuku,Cuku2}.
Their  behaviour can be interpreted roughly on the
lines introduced by Bouchaud \cite{Bo}.

\section{Geometry of phase-space}

As a first approach to the dynamical problem it is convenient to have
an understanding of the low-temperature structure of phase-space.
 We have at hand
two tools: the replica solution  and
the
Thouless-Anderson-Palmer (TAP) approach. The former gives us very detailed
information of the lowest-lying (`equilibrium') states which
contribute to the Gibbs-Boltzmann measure,
while the latter
 gives us an idea of what the landscape looks like for higher free-energies.
  Both approaches reveal that the two models we are considering have some
important qualitative differences, as the dynamics  also shows.

In the case of the $p$-spin spherical model, one level of replica symmetry
breaking is exact. The equilibrium states are separated by $O(N)$ barriers
(the difference in energy and free-energy
between them being $O(1)$) \cite{Kupavi}. The `size' of the
equilibrium states
is $q_{EA}$ while the overlap between different states is
$q_0=0$  \cite{Crso}.

The TAP equations for this model have solutions for a free-energy range
$(F_{1rsb}$, $F_{thres})$, corresponding
to an energy range $(E_{1rsb}$, $E_{thres})$.
The threshold energy
$E_{thres}$ (free-energy $F_{thres}$)
is greater than the equilibrium energy
 $E_{1rsb}$ (free-energy $F_{1rsb}$)
by a (temperature and $p$ dependent) $O(N)$ value.
Under the threshold the local minima of the TAP free-energy are separated by
$O(N)$ barriers while above the threshold there are no minima at all
\cite{Kupavi}. The free-energy Hessian of a solution has
eigenvalues larger than a certain $\lambda_{min}$ that
depends on the energy of the solution. For sub-threshold solutions
$\lambda_{min}$ is positive-definite, its value decreases with increasing
free-energy until it vanishes at the threshold.
As the temperature is changed the solutions
neither merge nor bifurcate, and their free energy
changes smoothly.

In the case of the SK model, the equilibrium states are also organised in an
ultrametric fashion, but with all possible distances between them allowed.
The barriers between equilibrium states are believed to be of only
$O(N^\alpha)$ with
$\alpha \sim 1/3$.
The solutions of the TAP equations tend to split as the temperature is lowered
in a second-order fashion. The Hessian of the solutions has a spectrum
going down to zero \cite{BM}, so that there is the possibility that
the barriers between neighbouring high free-energy TAP solutions
be finite.
For this  model
there seems to be no high free-energy
threshold below which all TAP solutions are separated
by infinite barriers and above which there are no solutions, as happens in
the  $p$-spin spherical model.

Fig. 1 shows schematically this structure of states
and barriers.

\section{Asymptotic out of equilibrium regime}

Let us now turn to the dynamics of an infinite system. We
take the thermodynamic limit $N \rightarrow \infty$ from the
outset
\footnote{ This is the main difference between
this  dynamical approach and the one proposed by Sompolinsky
\cite{So}.},
large times henceforth
 mean $t \rightarrow \infty$ {\it after}
$N \rightarrow \infty$. Under this hypothesis
there may be unsurmountable barriers.
We  consider a process in which
the system evolves from an initial configuration chosen at random.

Taking into account the structure of the phase-space, three possibilities
for the long-time dynamics can be expected:

\vspace{.3cm}

A. The system reaches the Gibbs-Boltzmann distribution in a characteristic time
   $t_{eq}$.

\vspace{.3cm}

B. The system equilibrates within a separate ergodic component. The
   distribution  is a
  Gibbs-Boltzmann distribution restricted to that sector of phase-space,
and it does not evolve after a characteristic time $t_{eq}$.

\vspace{.3cm}

C. The system does not reach a time-independent
 Gibbs-Boltzmann distribution within
{\em any fixed  sector} of phase-space at any time. There is no $t_{eq}$
such that for $t>t_{eq}$ the distribution has stabilised.

\vspace{.3cm}

Case A corresponds to an ordinary equilibration process, thermodynamical and
long-time dynamical calculations coincide. The equilibrium theorems hold
for times larger than $t_{eq}$.

Case B corresponds to falling in a high-lying stable state and a purely
statical
calculation does not necessarily yield the right values for order parameters
and free-energy. However, from the dynamical point of view this situation
is not very different from the situation  A: after the time $t_{eq}$
the system is  for all practical (dynamical) purposes as if in equilibrium.
Time-homogeneity and the fluctuation-dissipation theorems (FDT)
hold for any two times $t_1>t_2>t_{eq}$.

Case C
is what we find in mean-field spin-glass models.
As the system evolves,
 the dynamical free-energy density decreases, and the system
finds an ever decreasing portion of the landscape where it can move.
While at high temperatures this process ultimately leads the system to
the equilibrium state, at low temperatures the geometry of the
region of phase-space available at a certain time becomes more and more
 complicated, and
 the system is slowed down and does not reach an equilibrium state
in finite times. However, it does not remain permanently confined in
any  finite region within which  it is in local
equilibrium.

For instance, in the scenario of Bouchaud \cite{Bo},
one considers a phase-space with  `traps' separated by finite
barriers with a wide distribution of lifetimes. Traps visited at long times
tend to have long  lifetimes, for purely probabilistic reasons.
The system does not reach the true states (those with infinite lifetime)
 in finite
times, hence the name `{\em weak} ergodicity breaking'.

Interestingly enough, once the thermodynamic limit has been taken
there are no external parameters in the
dynamics of the system that go to infinity. Instead it is the age $t_w$
 (the time
elapsed since the  quench from above the transition
temperature) that regulates the slowing down of the system.
 For longer waiting times, the system has the opportunity
of finding deeper `traps' and narrower
channels, it becomes
 less susceptible to changes and thus it ages.

\vspace{1cm}

\section{Assumptions on the Correlation and Response Functions}

The solution of the models is based on some assumptions
that take into account the preceding scenario. We
now describe these assumptions.

\vspace{.3cm}

\noindent {\it i.} `Weak-ergodicity breaking'

\vspace{.3cm}

After any time $t_w$ (that may be interpreted as a waiting time)
the system continues to drift away and it reaches,
asymptotically, the maximum distance
compatible with the remanent magnetization.
Thus, the  correlation function
$C(t+t_w,t_w) \equiv (1/N) \,
\sum_{i=1}^N \overline {\langle s_i(t+t_w) s_i(t_w) \rangle} $ satisfies
\beq
\frac{\partial C(t+t_w,t_w)}{\partial t } \leq 0
\;\;\;\;\;\;\;\;\;
\frac{\partial C(t,t')}{\partial t' } \geq 0
\label{web}
\eeq
$t > t'$, and in the absence of a magnetic field
\beq
\lim_{t\rightarrow\infty} C(t+t_w,t_w) = 0 \;\;\;\;\; \forall \mbox{ fixed }
t_w
\label{limitc}
\eeq

This hypothesis is supported by the numerical simulations
of various mean-field  \cite{Cukuri,Strisc,Bacukupa} and realistic
\cite{Ri} models.
The numerical solution of the mean-field
dynamical equations
for the $p$-spin spherical model (see Ref. \cite{Cuku}) also support
this assumption:
in Fig. 2.a. we plot the decay of the correlation function
$C(t+t_w,t_w)$ {\it vs.} $t+t_w$. It is clear that the sign of the derivative
is negative for the whole time-interval plotted. Moreover, Fig. 2.b.
shows the decay of the correlation $C(t,t')$ as a funtion of $t'$ and
the derivative is positive in this case. ($T=.3$, $p=3$
in both cases.)
\footnote{We present these figures
to  show the qualitative tendency of the correlation and
response functions, a more precise
numerical analysis is in order to obtain quantitative results.}

\vspace{.3cm}

\noindent {\it ii.} `Weak long-term memory'

\vspace{.3cm}

The response to a constant (small) magnetic field applied from $t' = 0$
up to $t' = t_w$,
$h_{t_w}(t') = h \theta(t_w - t') \theta(t')$, {\it i.e.} the thermoremanent
magnetization, decays to zero for a long-enough time $t$
\beq
\lim_{t\rightarrow\infty} m_{t_w}(t) =
\lim_{t\rightarrow\infty} \int_0^{t_w} dt' G(t,t') = 0
\;\;\;\;\; \forall \mbox{ fixed } t_w
\eeq
The associated experimental curves \cite{Lusvnobe,Vihaoc}
and the simulations of the $D$-dimensional spin-glass cell
\cite{Cukuri} are compatible with this assumption.

It is this property that makes solutions to the long-time dynamics possible:
Because the memory of the system of its history is
strong when integrated over long times but weak at each given time,
one does not need
to solve all the short time (far off-equilibrium) details.

\vspace{.3cm}

\noindent {\it iii.}
Fast and slow relaxations.

\vspace{.3cm}

After a (long) time $t'$ there is a quick relaxation in a further time
$\tau =t-t'$ ($\tau$ short compared to $t'$)
 to some value $q$ (the dynamical
Edwards-Anderson parameter), followed by a
slower drift away.
Fast and slow relaxations correspond to relaxations within and away from
a trap, respectively.
Within a trap the system behaves as if it
were in a local equilibrium, the equilibrium
properties (Fluctuation-Dissipation Theorem
(FDT)
and homogeneity in time) are assumed to hold.
The value $q$ is the size of the traps
or the width of the channels
through which the system evolves.

\beqn
\begin{array}{rcl}
C(t,t') &=& C_{FDT}(t-t') + \cl(t,t')
\nn
& &
\nn
G(t,t') &=& G_{FDT}(t-t') + \gl(t,t')
\end{array} &
 \;\;\;\; G_{FDT}(t-t') = \frac{\partial C_{FDT}(t-t')}{\partial t'}
\eeqn
with
\beqn
\begin{array}{rclrcl}
C_{FDT}(0) &=& 1 - q
&
\lim_{t-t' \rightarrow\infty} C_{FDT}(t-t')
&=&
0
\nn
\cl(t,t) &=& q
&
\;\;\;\;\; \lim_{t\rightarrow\infty} \cl(t,t')
&=&
0
\; .
\end{array}
\eeqn
The relaxation away from a trap is slow:
\beq
\frac{\partial\cl(t,t')}{\partial t} \sim 0 \;\;\;\;\;\;\;
 \frac{\partial\cl(t,t')}{\partial t'} \sim 0 \; \; \;
\mbox{ for large }\;\;t,t'
\label{slow}
\eeq

In the simulations of the  $D$-dimensional hypercubic cell of Ref.
\cite{Cukuri},
the $\log(C(t+t_w,t_w))$ vs. $\log(t)$ curves
show that for short times $t=(t+t_w)-t_w$ compared to $t+t_w$
all the curves for different waiting times $t_w$ lie on each other. Hence,
the correlation function is homogeneous in time for that range. The fast
relaxation to a value $q$ can be checked in the curves presented in
Ref. \cite{Cuku2}.

In  the numerical solution of the mean-field
dynamical equations of the $p$-spin spherical model we also find
support for this hypothesis. The Figs. 2.a and 2.b show a `fast' decay
from the value $1$ at equal times to a certain value.
This is more clear in Fig. 2.b. where $C(t,t')$ decays fastly from $1$ to
a value $q \sim .9$ This is justified since the total time involved
is larger
and the interval for the FDT decay must be short compared to the total
time but it itself must be long.

\vspace{.3cm}

Two further properties that are  in part consequences of the assumptions
already made are:

\vspace{.3cm}

\noindent {\it iv.}
For long times $t,t'$
${\cal C}$ and  ${\cal G}$ are related by

\beq
{\cal G}(t,t')= X[{\cal C}(t,t')] \,
\frac{\partial {\cal C}(t,t')}{\partial t'} \; \theta(t-t')
\label{xeq}
\eeq
where $X$ depends on the times {\em only through} ${\cal C}$.

\vspace{.3cm}

The relation (\ref{xeq})
with $X[t,t']= \hat X[t, {\cal C}(t,t')]$ implies no assumption.
 One can show
that in the large times limit the explicit dependence on $t$
can be neglected if the energy, the susceptibility, and all the
`generalised susceptibilities'  \cite{Cuku,Cuku2} have a limit.

If we supplement the definition of $X[z]$
with $X[z]=1$ for $q<z<1$
then the relation (\ref{xeq}) holds for all $C(t,t')$, $t'$ large.
$X[z]$ may be discontinuous in $z=q$, where it
jumps from $X[q]$ to $1$ ($X=1 \Leftrightarrow $ FDT).

An assumption concerning the function $X(C)$, suggested by the simulations
is:

\vspace{.3cm}

\noindent {\it v.} $X[\cl]$ is an increasing function of $\cl$.

\vspace{.3cm}

\noindent {\it vi.}
`Triangle relations' between the correlations at three large times
$t_{min} \leq t_{int} \leq t_{max}$

\vspace{.3cm}

The monotonicity
of the correlation function with respect to both times
described above, allows to write
\beq
C(t_{max},t_{min})= f[C(t_{max},t_{int}),C(t_{int},t_{min}),t_{min}]
\eeq

We can consider the limit for large $t_{min}$, and for fixed
$C(t_{max},t_{min})$. The fact that such a limit exists
implies that we can write, for large enough $t_{min}$:
\beq
C(t_{max},t_{min})= f[C(t_{max},t_{int}),C(t_{int},t_{min})]
\label{triangle}
\eeq
where the explicit dependence on times have dissapeared,
and the three correlations are related by `triangle' relations.
It is convenient to define the inverse relation:
\beq
C(t_{max},t_{int})=\overline f[C(t_{max},t_{min}),C(t_{min},t_{int})]
\eeq

A simple computation involving four times shows that the function
$f$ is {\em associative} \cite{Cuku2}
\beq
f(f(q_1,q_2),q_3)=f(q_1,f(q_2,q_3))
\eeq
The property of weak ergodicity breaking implies that
\beq
f(q_1,q_2) \leq \min (q_1,q_2)
\label{web1}
\eeq
One can now classify all the possible associative $f$ that satisfy
(\ref{web1}) as follows: consider the values $q_i^*$ for which
\beq
f(q_i^*,q_i^*)=q_i^*
\; .
\eeq
We call them fixed points. Fixed points and
intervals between consecutive fixed points are `correlation scales'.
The latter we call `discrete scales' and, in
particular, the interval $(q,1)$ is one of them.
A dense set of $q_i^*$ corresponds to a dense set of scales.

It is shown in Ref. \cite{Cuku2} that if $q_1$ and $q_2$
belong to different scales then
\beq
f(q_1,q_2) = \min (q_1,q_2)
\label{ultra}
\eeq
If, instead, they belong to the same discrete scale:
\beq
f(q_1,q_2) = \jmath_k^{-1} [\jmath_k(q_1)\jmath_k(q_2)]
\; ,
\label{scala}
\eeq
where $\jmath_k$ are functions that may be different for each
discrete scale.

In Fig. 3 we sketch the function $f(a,a)$ for a ferromagnet, a
paramagnet, the $p$-spin spherical spin-glass and the SK spin-glass.
In the next sections we present the analysis of the $p$-spin spherical
model and we discuss the corresponding results for ths SK model
\cite{Cuku2}.

\section{A Simple Example}

The analysis and assumptions in the preceding section are quite general
and model-independent.
The precise form of $X(C)$, the limits of the correlation scales,
and the functions $\jmath_k$ associated with the discrete scales (if any)
have to be extracted from the dynamical equations of each model.

The analysis of the SK model along these lines was done in Ref. \cite{Cuku2},
here we work out the dynamics of the $p$-spin spherical
model \cite{Cuku} within this formalism.
We shall afterwards discuss the differences between these two representative
models.

The mean-field equations of motion for the $p$-spin spherical
model read:
\beqn
\frac{\partial C(t,t')}{\partial t}
&=&
- \, (1- p \beta \,  {\cal E}(t)) \, C(t,t')
+
2 \, G(t',t)
\nn
& &
+
\,
\mu \int_0^{t'} dt'' \, C^{p-1}(t,t'') \, G(t',t'')
\nn
& &
+
\,
\mu \, (p-1) \int_0^t dt'' \, G(t,t'') \, C^{p-2}(t,t'') \, C(t'',t')
\; ,
\label{corr}
\\
\frac{\partial G(t,t')}{\partial t}
&=&
- \,(1- p \beta \, {\cal E}(t)) \, G(t,t')
+ \delta(t-t')
\nn
& &
+
\,
\mu \, (p-1) \int_{t'}^t dt'' \, G(t,t'') \, C^{p-2}(t,t'') \, G(t'',t')
\; ,
\label{resp}
\eeqn
where $\mu = p \beta^2/2$ and the energy ${\cal E}$:
\beq
{\cal E}(t)=
-\mu
\int_0^t dt'' \, C^{p-1}(t,t'') \, G(t,t'')
\label{int}
\eeq

Considering assumption {\it iii} we can separate the small time difference
(FDT) regime from the regime of widely separated times.
We hence obtain:
\beq
(\frac{\partial}{\partial \tau} + 1)
C_{\rm FDT}(\tau)
+
(\mu + p \beta \, {\cal E}_\infty)
\,
(1 - C_{\rm FDT}(\tau))
=
\mu \int_0^\tau d\tau''
\,
C^{p-1}_{\rm FDT}(\tau-\tau'') \,
\frac{dC_{\rm FDT}}{d\tau''}(\tau'')\nonumber \\
\label{sompoz}
\eeq
with the asymptotic energy ${\cal E}_\infty$
given by
\beq
{\cal E}_\infty
=
-
\frac{\beta}{2} \, \left[
(1-q^p)
+
p
\int_0^t dt'' \, {\cal G}(t,t'') \, {\cal C}^{p-1}(t,t'')
\right]
\label{eninf}
\; .
\eeq
The correlation decays to a value $q$ determined by
\beq
-(1- p \beta \, {\cal E}_\infty) + \mu \, (1-q^{p-1}) = -\frac{1}{1-q}
\label{qFDT}
\; .
\eeq
Equation (\ref{sompoz}) corresponds to the dynamics \`a la
Sompolinsky-Zippelius of
the relaxation `within a trap'  \cite{CHS}.

For the long time differences we obtain:
\beqn
0
&=&
{\cal G}(t,t')
\left [
-(1-q)^{-1}
+
\mu \, (1-q) (p-1) \, \, {\cal C}^{p-2}(t,t')
\right ]
\nn
& &
+
\mu \, (p-1) \,
\int_0^t
dt'' \, {\cal G}(t,t'')
\, {\cal C}^{p-2}(t,t'')
\, {\cal G}(t'',t')
\; ,
\label{gcursiva}
\\
0
&=&
{\cal C}(t,t')
\left [
-(1-q)^{-1} + \mu \, (1-q) \,  \, {\cal C}^{p-2}(t,t')
\right ]
\nn
& &
+
\mu \,
\int_0^{t'}
dt'' \, {\cal C}^{p-1}(t,t'') \,
{\cal G} (t',t'')
\nn
& &
+
\mu \, (p-1) \,
\int_0^t
dt'' \, {\cal G}(t,t'') \, {\cal C}^{p-2}(t,t'') \,
{\cal C} (t'',t')
\label{ccursiva}
\; .
\eeqn

We have now two (coupled) sets of equations, corresponding to two time regimes.
It is important to note that in order to claim that eqs.
(\ref{gcursiva}) and (\ref{ccursiva})
are asymptotically valid {\em we need
the property of weakness of the long-term memory}, which allows us
to disregard in the integrals any finite-time interval.
We can then neglect the fact that for the initial times the asymptotic
equations do not hold.

Above the critical temperature ${\cal G} \sim 0$, the long-term
memory is absent, and we are left with
only equation (\ref{sompoz}) \cite{CHS}.
This is the same equation that
arises in the mode-coupling theory of structural glasses \cite{Go}.

In the glass phase, in order to determine any quantity,
 we have to solve the whole set of eqs. (\ref{sompoz}),
(\ref{gcursiva}) and (\ref{ccursiva}).

We concentrate here on the solution
of  eqs. (\ref{gcursiva})
and (\ref{ccursiva}).
 We have neglected in  them the time derivatives following  {\it iii}.
This brings as a consequence the fact that these equations (but not
eq. (\ref{sompoz})) have a continuous set of reparametrization invariances.
Indeed, from a solution ${\cal C}$, ${\cal G}$ we obtain infinitely
many others:
\beq
{\hat{\cal C}}({\hat t},{\hat t'})={\cal C}(h(t),h(t')) \; , \;\;\;\;\;\;\;
{\hat {\cal G}}({\hat t},{\hat t'})= h'(t') \, {\cal G}(h(t),h(t'))
\; ,
\label{reparam1}
\eeq
with $h$ any increasing function.
This invariance is a consequence of having neglected the time
derivatives in making the asymptotic limit. The full
dynamical equations have no such invariances; because of causality
their solution is {\em unique}.
We face the {\em selection problem}, a rather common phenomenon in
the asymptotic limit of solutions of differential equations.
In this work we discuss solutions only modulo reparametrizations, the
selection problem for this kind of dynamics remains to be solved.

Let us now use the assumptions {\it iv}  and  {\it vi}
to write (\ref{gcursiva})
and (\ref{ccursiva}) in terms of $X$ and $\bar f$.
Defining
\beq
F[\cl] \equiv - \int_{\cal C}^q d \cl' \, X[\cl'] \;\;\;\;\;\;\;
H[\cl] \equiv -\int_\cl^q d\cl' \, \cl'^{p-2} \, X[\cl']
\label{Fdef}
\eeq
the dynamical equations (\ref{gcursiva})
and (\ref{ccursiva})  become
\beqn
0
&=&
-(1-q)^{-1} \, F[\cl]
+
\mu \, (p-1) \, (1-q) \, H[\cl]
\nn
& &
+
\mu \, (p-1) \, \int_\cl^q d \cl' \, X[\cl'] \, \cl'^{p-2} \,
F[ \overline f(\cl', \cl)]
\; ,
\label{dync1}
\\
0
&=&
\left[
-(1-q)^{-1} + \mu \, (1-q) \, \cl^{p-2}
\right]
\, \cl
-
\mu \, (p-1)  \,
\int_0^\cl d\cl' \, \cl'^{p-2} \,
F[\overline f(\cl, \cl')]
\nn
& &
+
\mu \, (p-1) \, \int_0^\cl d\cl' \, X[\cl'] \, \cl'^{p-2} \,
\overline f(\cl, \cl')
\nn
& &
+
\mu \, (p-1) \, \int_{\cal C}^q d\cl' \, X[\cl'] \, \cl'^{p-2} \,
\overline f(\cl', {\cal C})
\; ,
\label{dyng1}
\\
{\cal E}(t)
&=&
-\frac{\beta}{2}
\left[
(1-q^p)
+
p \, \int_0^q   d\cl' \, X[\cl'] \, \cl'^{p-1}
\right]
\; .
\label{dynz}
\eeqn

In using the relation (\ref{xeq}) and (\ref{triangle})
we have eliminated the times. Hence we have {\em divided by the
reparametrization group}. Indeed,  (\ref{xeq}) and (\ref{triangle})
are reparametrization-invariant relations themselves.

The next step is to investigate the possibility of having
`discrete' correlation scales. The method is to propose that there is a scale,
and then check
which limits the equations allow the scale to have.
Using the properties of the function $\overline f$ described in the previous
section \cite{Cuku2}, these equations can be rewritten
within a discrete scale $ \cl \in (a_2^*, a_1^*)$, $0 < a_1^* < q$,
\beqn
0
&=&
-(1-q)^{-1} \, F[\cl]
+ \mu \, (p-1) \, (1-q) \, H[\cl]
- \mu \, (p-1) \, F[\cl] \, H[a_1^*]
\nn
& &
+
\mu \, (p-1) \, \int_\cl^{a_1^*} d\cl' \, X[\cl'] \, \cl'^{p-2} \,
F[\overline f(\cl',\cl)]
\; ,
\label{dyng2}
\\
0
&=&
\left[
-(1-q)^{-1} + \mu \, (1-q) \, \cl^{p-2}
- \mu \, (p-1) \, H[a_1^*]
\right]
\,
\cl
\nn
& &
+
\mu \, (p-1) \int_0^{a_2^*} d\cl' \,
\cl'^{p-2} \,
\left(
\cl' X[\cl'] - F[\cl']
\right)
\nn
& &
+
\mu \, (p-1) \, \int_{a_2^*}^\cl d\cl' \, \cl'^{p-2} \,
\left(
X[\cl'] \overline f(\cl,\cl') -  F[\overline f(\cl, \cl')]
\right)
\nn
& &
+ \mu \, (p-1) \,
\int_{\cl}^{a_1^*} d\cl' \, X[\cl'] \, \cl'^{p-2} \, \overline f(\cl', \cl)
\; .
\label{dync2}
\eeqn
Evaluating eq.(\ref{dyng2}) in $\cl = a_1^*$ we have
\beq
-(1-q)^{-1} F[a_1^*] + \mu (p-1) \,
\left(
(1-q) - F[a_1^*]
\right)
\,
H[a_1^*] = 0
\; ,
\label{eq1}
\eeq
and differentiating  eq.(\ref{dyng2}) w.r.t. $\cl$ and evaluating in
$\cl = a_1^*$ we get
\beq
(1-q)
\left(
{a_1^*}^{p-2} - q^{p-2}
\right)
- {a_1^*}^{p-2} F[a_1^*] - H[a_1^*] = 0
\; ,
\label{eq2}
\eeq
provided $X[a_1^*] \neq 0$. From eqs. (\ref{eq1}) and (\ref{eq2})
$F[a_1^*]$ is given by
\beq
F[a_1^*]
=
(1-q)
\left(
1 - \sqrt{ \frac { q^{p-2}}{{a_1^*}^{p-2}} }
\right)
\; .
\label{feq}
\eeq
Differentiating again eq. (\ref{dyng2})  w.r.t. $\cl$, evaluating the result
in $a_1^*$ and using the previous eqs. (\ref{eq2}) and (\ref{feq})
we finally get
\beq
X[a_1^*]
=
\frac{ (p-2)(1-q)}{a_1^*} \sqrt{ \frac{ q^{p-2} }{ {a_1^*}^{p-2} } }
\; .
\eeq
Since we assume that $X[\cl]$ is an increasing function of $\cl$
(assumption {\it v}) then
\beq
a_1^* = q
\eeq
and
\beq
X[a_1^*] = X[q] = \frac{(p-2)(1-q)}{q}
\; .
\eeq

Once we have shown that the upper limit of the discrete scale must be
$q$ we need to obtain its lower limit. First, we shall show
that $X[\cl]$ must be constant inside the scale, {\it i.e.} $X[\cl] = X =
(p-2)(1-q)/q$, $\forall \cl \in (a_2^*, q)$.
Computing the derivatives of eqs. (\ref{dyng2}) and (\ref{dync2})
w.r.t. $\cl$, multiplying the former by $X[\cl]$ and subtracting we have
\beqn
0
&=&
\int_{a_2^*}^\cl d\cl' \, \cl'^{p-2} \, X[\cl]
\,
\frac{ \overline f(\cl, \cl')} {\partial \cl}
\,
\left(
X[\overline f(\cl,\cl')] - X[\cl']
\right)
\nn
& &
+
\int_\cl^q d\cl' \, \cl'^{p-2} \, X[\cl']
\,
\frac{ \overline f(\cl', \cl)} {\partial \cl}
\,
\left(
X[\overline f(\cl',\cl)] - X[\cl]
\right)
\; .
\eeqn
This equation has $X[\cl]=X$ as a solution. To show that it is the only
admisible solution we derivate again w.r.t.
$\cl$ and evaluate in $\cl = a_1^*=q$ to get
\beq
\frac{d X[\cl']}{d\cl'} = 0 \;\;\; \Rightarrow \;\;\; X[\cl'] = X
\; ,
\eeq
$\forall \cl \in (a_2^*,q)$.

Finally, using the constancy of $X$ we calculate $F[\cl]$ and $H[\cl]$
\beq
F[\cl] = -X(q-\cl)
\;\;\;\;\;\;
H[\cl] = -\frac{X}{p-1} (q^{p-1} - \cl^{p-1})
\; ,
\eeq
and from eq. (\ref{Fdef})  we get
\beq
a_2^* = 0 \;\;\; \mbox{or} \;\;\; a_2^* = q
\; .
\eeq
Hence the discrete scale is `empty', there are no discrete scales at all
but a continuos set of fixed points from 0 to q,
or there only one discrete scale that spans the whole interval
$(0,q)$ (apart from the FDT sector). We have checked that the former
possibility is excluded by the dynamical equations, namely (\ref{dyng1}).

We can now turn to the explicit solution within the discrete scale
spanning the interval $(a_2^*=0,a_1^*=q)$.
Eqs. (\ref{dyng2}) and (\ref{dync2}) can now be written as
\beqn
0
&=&
q - \cl - \mu \, (p-1) \, (1-q)^2 \, (q^{p-1}-\cl^{p-1})
\nn
& &
+
\mu \, (p-1) \, (1-q) \, X \,
\int_\cl^q d\cl' \, {\cl'}^{p-2} \, \overline f(\cl',\cl)
\; ,
\label{eq11}
\\
0
&=&
\cl
\left[
- 1 + \mu \, (p-1) \, (1-q)^2 \cl^{p-2}
\right]
\nn
& &
+
\mu \, (p-1) \, (1-q) \, X \,
\int_\cl^q d\cl' \, {\cl'}^{p-2} \, \overline f(\cl',\cl)
\; .
\label{eq12}
\eeqn
The value $q$ can be
immediately determined from these equations; they imply
\beq
\frac{1}{p-1} = \mu q^{p-2} (1-q)^2
\; .
\eeq
{}From the analysis presented in Ref. \cite{Cuku2} we know that the
solution within a discrete scale has the form
\beq
\overline f(\cl', \cl)
=
\jmath^{-1} \left( \frac{\jmath(\cl)}{\jmath(\cl')}
\right)
\; ,
\eeq
$\cl < \cl'$.  We should now use the equations (\ref{eq11}) and (\ref{eq12})
to determine the function $\overline f(\cl', \cl)$.

Making the change of variables:
\beq
\rho \equiv \ln \left(\frac{{\cal C}}{q} \right)
\;\;\;\; ; \;\;\;\; k(\rho) \equiv \jmath^{-1}(q e^{\rho})
\eeq
eqs. (\ref{eq11}) and (\ref{eq12}) both become
\footnote{Remarkably, this is the same equation that one obtains for long
times in the Sompolinsky (time-homogeneous) dynamics \cite{CHS} with
$\rho$ playing the role of time-differences.}
\beq
k(\rho) \,  [ 1-k^{p-2}(\rho)] \, + \,
\frac{p-2}{p-1} \, \int_{0}^{\rho} d \rho'
\,
\frac{\partial k^{p-1}(\rho')}  {\partial \rho'} \, k(\rho - \rho')=0
\; .
\eeq
 The smooth solutions to this equation are:
\beq
k(\rho)=e^{\alpha \rho} \;\;\;\;\;
\Rightarrow \;\;\;\;\; \overline f(\cl', \cl)=
q^{-1} \, \frac{\cl'}{ \cl}
\eeq
Which in turn implies \cite{Cuku2} that ${\cal C}$ is of the form:
\beq
{\cal C}(t,t')=q \, \frac{h(t')}{h(t)}
\eeq
for any increasing $h$. {\em This is as far as we can go analytically
without
solving the selection problem}. The numerical solutions suggest  \cite{Cuku}
that $h(t)=t^{\gamma}$.

For large $t_w$ one has, as $t-t_w \rightarrow \infty$
\beq
\cl(t+t_w,t_w) \simeq q \left( \frac{h(t_w)}{h(t)} \right)
\eeq
which explicitly shows aging.

\section{Discussion}

We have shown how to obtain  some
analytical results for  spin-glass mean-field dynamics,
on the basis of the assumptions made.
In this  solution the aging phenomena are explicit.
The solution is not complete, some quantities we  have only obtained
{\em modulo time-reparametrizations}. However, the long-time
limits of all quantities that depend on a single
time (energy, magnetization, $q$, etc) are unaffected by reparametrizations,
and hence fully determined.

We have found a solution for the p-spin spherical model
with one discrete scale
apart from the FDT scale.
Let us now briefly describe what happens when one applies exactly the
same procedure to the SK model \cite{Cuku2}.
 It is found there that the dynamical equations
do not admit any discrete scales (apart from the FDT $(q,1)$ scale),
but only a dense set of scales in the
correlation interval $(0,q)$.
This is very much like in the static treatment, discrete scales playing the
same role as levels of replica symmetry breaking.
There is however a difference:
while for the SK model one
finds that the values of the asymptotic energy,  $q$, and transition
temperature coincide (to  $O(N))$
with those obtained in the static treatment, for the $p$-spin model
they do not.
Indeed, for this model one finds that the values for
$q$ and $E$ correspond to
those of the threshold level.

Another difference between the two models is in the behaviour of the
thermoremanent magnetization. In the $p$-spin spherical model it decays
with a rate that is inversely proportional to the waiting time the field has
been on. In the SK model the decay of the magnetization after a long
waiting time
can be seen as taking place in steps, each of which
is takes much longer  than the preceding one. Hence, the
SK model has a larger degree of `freezing' of the magnetization.

To conclude, let us remark that the dynamical spin-glass phase can
be viewed as a phase where a symmetry is spontaneously broken.
The dynamical process
has a supersymmetry (SUSY) group of invariances \cite{Frku}
with three generators
\beq
{\bf D'}\;\;\; ; \;\;\; {\overline {\bf D'}} \;\;\; ; \;\;\;
[{\overline {\bf D'}} , {\bf D'}]_+= \frac{\partial }{\partial t_1}+
\frac{\partial }{\partial t_2}
\eeq
associated with probability conservation, FDT and time-homogeneity.
The two-time  $(t,t')$ functions are defined in the time region
whose two boundaries are  the lines  $t=0$ and $t'=0$, respectively.

The full SUSY group is broken down in certain time-regions
down to the subgroup generated by only ${\bf D'}$.
In the high temperature phase the effect of the initial (boundary)
condition is to break the SUSY in a region of width $\simeq t_{eq}$
near the boundaries. In the low temperature phase the SUSY-breaking
persists in an infinite region away from the boundaries.
This is much like the effect of symmetry-breaking boundary conditions
in  ordinary sytems.

\pagebreak

\end{document}